\documentclass[twocolomn,fleqn]{pasj00}
\usepackage{longtable}
\usepackage{lscape}

\draft

\def\commenta{$^*$}
\def\commentb{$^\dagger$}

\newcounter{author}
\def\authorcount#1#2{\refstepcounter{author}\label{#1}
                     \altaffiltext{\ref{#1}}{#2}}

\begin{document}
\SetRunningHead{Pavlenko et al.}{EZ Lyncis Second Visit to Instability Strip}

\Received{201X/XX/XX}
\Accepted{201X/XX/XX}

\title{Dwarf Nova EZ Lyncis Second Visit to Instability Strip}

\author{Elena~P.~\textsc{Pavlenko},\altaffilmark{\ref{affil:CrAO}*}$^,$\altaffilmark{\ref{affil:Kyoto}}
        Taichi~\textsc{Kato},\altaffilmark{\ref{affil:Kyoto}}

        Aleksei~A.~\textsc{Sosnovskij},\altaffilmark{\ref{affil:CrAO}}
        Maksim~V.~\textsc{Andreev},\altaffilmark{\ref{affil:Terskol}}$^,$\altaffilmark{\ref{affil:ICUkraine}}
        Tomohito~\textsc{Ohshima},\altaffilmark{\ref{affil:Kyoto}}
        Aleksander~S.~\textsc{Sklyanov},\altaffilmark{\ref{affil:Kazan}}
        Ilfan~F~.\textsc{Bikmaev},\altaffilmark{\ref{affil:Kazan}}
        Almaz~I.~\textsc{Galeev},\altaffilmark{\ref{affil:Kazan}}
}

\authorcount{affil:CrAO}{
     Crimean Astrophysical Observatory, 98409, Nauchny 19/17, Crimea, Ukraine}

\authorcount{affil:Kyoto}{
     Department of Astronomy, Kyoto University, Kyoto 606-8502}
\email{$^*$eppavlenko@gmail.com}

\authorcount{affil:Terskol}{
     Institute of Astronomy, Russian Academy of Sciences, 361605 Peak Terskol,
     Kabardino-Balkaria, Russia}

\authorcount{affil:ICUkraine}{
     International Center for Astronomical, Medical and Ecological Research
     of NASU, Ukraine 27 Akademika Zabolotnoho Str. 03680 Kyiv,
     Ukraine}

\authorcount{affil:Kazan}{Kazan Federal University, Kremlevskaya str.18,
     Kazan 420008, Russia
     }



\maketitle

\begin{abstract}
The analysis of 14 periodograms of EZ Lyn for the data spaced
over 565~d in 2012--2014 (2--3.5~yr after 2010 outburst) yielded
the existence of the stable signals around 100 c/d and three
signals around 310~c/d, 338~c/d and 368~c/d (the corresponding periods
are 864~s, 279~s, 256~s and 235~s). We interpret them
as independent non-radial pulsations of the white dwarf in
EZ Lyn, but a possibility that a linear combination of frequency
at 100~c/d and harmonic of orbital period could produce the frequency
at 368~c/d also cannot be excluded.
The signal at 100~c/d was detected during the first stay in
the instability strip as a transient one. The period at 338~c/d,
is a known non-radial pulsation EZ Lyn entered
the instability strip after the 2010 outburst. We detected
the signals around 310~c/d and 368~c/d for the first time.
We applied the two-dimensional least absolute shrinkage and
selection operator (Lasso) analysis for the first time
to explore  the behavior of these signals on the scale of hours for
nightly runs of observations having duration of 6--12~hr.
The Lasso analysis revealed the simultaneous existence of
all three frequencies (310~c/d, 338~c/d and 368~c/d)
for majority of nights of observations, but with variable amplitudes
and variable drifts of frequencies by 2--6 percents on a time scale
of $\sim$5--7 hr. The largest drift we detected corresponded
to 17.5~s in period in $\sim$5 hours.
\end{abstract}

\section{Introduction}

Cataclysmic variable (CV) is a close binary system
containing a late-type component transferring matter via
the inner Lagrangian point onto the primary component that
is white dwarf.  Due to an angular momentum loss,
the CV with hydrogen-dominant atmosphere evolves to
an orbital period minimum of 76.2~min \citep{kni06CVsecondary}
with decreasing of mass transfer rate up to
$10^{-11}$ solar mass/year [\citet{how95TOAD};
\citet{kol99CVperiodminimum}].
The last circumstance enables the white dwarf to dominate
the total radiation of CV, so such short-period systems are
attractive for studies of the white dwarf,
in particular to search for the white dwarf pulsations.
Indeed, since 1998 non-radial pulsations of 14 accreting
white dwarfs were found among the cataclysmic variables.
They are GW Lib \citep{war98gwlibproc},
PQ And \citep{van05pqand}; V455 And \citep{ara05v455and};
SDSS J133941.11$+$484727.5 = V355 UMa \citep{gan06j1339};
SDSS J091945.11$+$085710.1 \citep{muk07j0745j0919};
SDSS J151413.72$+$454911.9 = PP Boo \citep{nil06SDSSCVpuls};
SDSS J074531.91$+$453829.5 = EQ Lyn \citep{muk07j0745j0919};
RE J1255$+$266 \citep{pat05j1255};
SDSS J150722.33$+$523039.8 = OV Boo (\cite{pat08j1507}; \cite{szk10CVWDpuls});
EZ Lyn \citep{pav09j0804WD};
SDSS J161033.64$-$010223.3 = V386 Ser \citep{wou04j1610};
SDSS J145758.21$+$514807.9 and BW Scl \citep{uth12j1457bwscl}.
With two exceptions (OV Boo and RE J1255$+$266), their orbital periods
are within the region of 80--90 min.

Knowledge of WD pulsations is important for measuring of
its stellar mass, core composition, age, rotation rate,
magnetic field strength, and distance
(\cite{win08pulsWDreview}; \cite{fon08pulsWDreview}).

EZ Lyn was discovered by \citet{szk06SDSSCV5} as
a short period (0.059~d) quiescent dwarf nova with
an underlying white dwarf.  \citet{pav07j0804} first found
it at the superoutburst.  The large amplitude and series of
11 rebrightenings implied that this star belongs to
the WZ Sge-type subclass of dwarf novae.
After the outburst, \citet{zha08j0804} found series of
mini-outbursts with an amplitude about 0.5 mag.

In 2010 EZ Lyn had experienced the next superoutburst, which
was discovered by H. Maehara \citep{Pdot3}.
Contrary to the 2006 superoutburst, the 2010 superoutburst
had six rebrightenings that is almost two times less than
in the 2006 superoutburst.  One month after the end
of the main outburst, EZ Lyn was $\sim$1 magnitude fainter in the
2010 outburst than in the 2006 one.

The superhump period of EZ Lyn is 0.060~d
[\citet{she07j0804}, \citet{pav07j0804}, \citet{kat09j0804}],
and its orbital period is 0.059005~d \citep{kat09j0804}.
\citet{kat09j0804} first clarified the grazing eclipse of
EZ Lyn during the 2006 superoutburst.

Eight months after the 2006 outburst, \citet{pav09j0804WD}
discovered the 756~s (12.6~min) brightness
modulation that was interpreted as the nonradial pulsation of
the white dwarf.
This pulsation has been regularly detected and displayed
a drift from 732~s to 768~s
on a scale of $\sim$900~d \citep{pav12ezlyn},
and this wandering did not depend on time.
Besides this stable dominating period, other periodic
signals have appeared
time by time between 846~s and 1302~s \citep{pav12ezlyn}.

Seven months after the 2010 outburst, EZ Lyn entered
the instability strip the second time,
but with a new 256--257~s pulsation
(\cite{pav12ezlyn}, \cite{szk13ezlyn}).
The drift of pulsation period as well as the change of
pulsation amplitudes appears to be a common feature of
accreting pulsators (\cite{uth12j1457bwscl}; \cite{muk10v386ser}).
However, it is not clear what could be
the shortest time scale of pulsation drift and
change of its amplitude.

Here, during the second stay of EZ Lyn in the instability strip
we present the detailed study of its periodic signals that
are shorter then orbital period.  We used least absolute shrinkage
and selection operator (Lasso) analysis
(\cite{lasso}; \cite{kat12perlasso}) for the first time
to study the behavior of such signals on a time scale of hours.

\section{Observations and data reduction}
Observations of EZ Lyn have been carried out
in the Crimean Astrophysical Observatory,
the Terskol Observatory and using the Russian-Turkish 1.5-m
telescope (RTT150) at TUBITAK National Observatory (Turkey)
in 2012--2014 during 14 nights.  The standard
aperture photometry (de-biasing, dark subtraction and flat-fielding)
was used for measuring of the EZ Lyn and comparison stars.
We used the comparison stars the same as \citet{zha08j0804}.
The observations have been carried out without usage of filters.
Typical accuracy for the Crimean and Terskol observations
was 0.005--0.017 mag and those of TUBITAK data was about 0.025 mag.
The log of observation is given in table \ref{tab:log}.
Before the analysis we subtracted the smooth long-time
trend for each night by using locally-weighted
polynomial regression (LOWESS, Cleveland 1979).
The times of observations are expressed in
Barycentric Julian Dates (BJD).
We used a phase dispersion minimization (PDM; \cite{PDM}), whose
1$\sigma$ errors was estimated by the methods of
\citet{fer89error} and \citet{Pdot2}; fast Fourier
transform (FFT) [ISDA package, \citet{pel80freqanalysis}].
The false-alarm probability (FAP) was estimated according to
\citet{sca82}.

Lasso was introduced for period analysis of variable stars
first time by \citet{kat12perlasso}, which yielded very sharp
signals.  A two-dimensional Lasso power spectrum was found
to be very effective in detecting varying multiple signals in
Kepler observations of CVs [\citet{kat13j1924};
\citet{osa13v344lyrv1504cyg}; \citet{kat13j1939v585lyrv516lyr}].
This method has been applied to the unevenly sampled
ground-based data (\cite{Pdot5}; \cite{ohs14eruma}), and
the resultant two-dimensional Lasso power spectra have been
proven to have high frequency resolutions and are less
affected by aliasing in unevenly sampled data than
the conventional Fourier analysis.

\begin{table}
\caption{Log of the observations}\label{tab:log}
\begin{center}
\begin{tabular}{cccc}
\hline
 Time  &Telescope/CCD\commenta & Exposure & $N$\commentb\\
BJD 2456000+ & &(s) & \\
\hline
236.360--236.619&A&60 &242 \\
237.341--237.488&A&45 &181 \\
244.518--244.626&A&60 &120 \\
245.336--245.623&A&45 &415 \\
247.340--247.627&A&60 &304 \\
248.388--248.627&A&90 &202 \\
249.333--249.488&B&30 &246 \\
250.427--250.661&B &20 &716 \\
306.238--306.579&B&30 &579 \\
307.156--307.672&B& 20 &1785 \\
385.208--385.259&B&15 &259 \\
697.184--697.676&B&30 &738 \\
711.407--711.547&C&60 &200\\
801.298--801.370&B&30 &159 \\
\hline
 \multicolumn{4}{l}{\commenta A: 2~m Terskol/FLI PL430;} \\
 \multicolumn{4}{l}{B: 2.6~m ZTSh/Apogee E47;}\\
 \multicolumn{4}{l}{C: 1.5~m RTT-150/TFOSC}\\
 \multicolumn{4}{l}{\commentb $N$: Number of images} \\

\end{tabular}
\end{center}
\end{table}

\section{Periods}

\begin{figure}
\begin{center}
\FigureFile(80mm,80mm){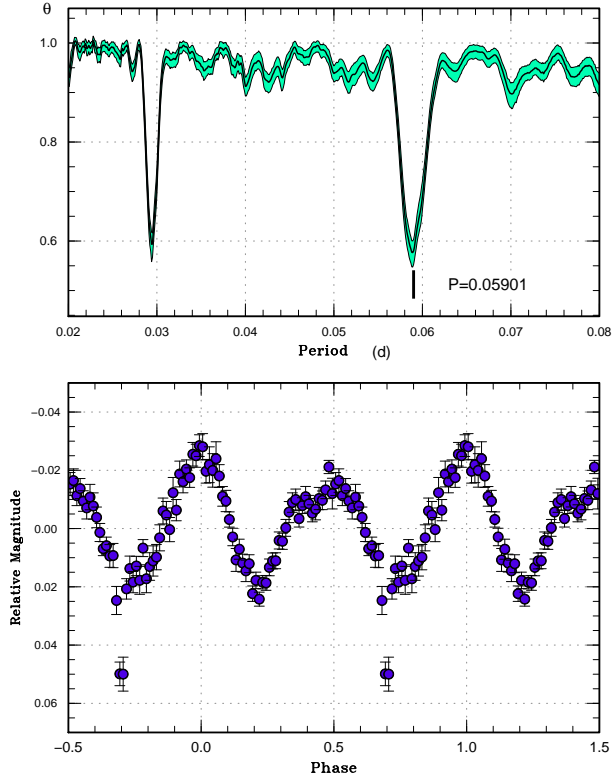}
\end{center}
\caption{Above: PDM for 2013 January 14 (BJD 2456307). Below: phase-averaged
data folded on the orbital 0.0590048~d period. The zero-epoch is BJD 2456307.16123. For clarity data are reproduced twice.}
\label{fig:14jan}
\end{figure}

The complexity in the analysis of potential pulsations of
the white dwarf in EZ Lyn is that they are contaminated by
the orbital modulation.  Both signals may be of variable amplitudes
[\citet{szk06SDSSCV5}, \citet{pav09j0804WD}],
but typically the orbital signal is the dominating one.  The PDM spectrum
and the mean orbital light curve
(affected by the less-amplitude pulsations) for the data of
2013 January 14 are presented in figure \ref{fig:14jan}.
The strong first harmonic of the orbital signal
is caused by the two-humped profile of the light curve.
Its average amplitude is 0.08 mag,
a grazing eclipse at the phase 0.7
is clearly seen.
For this pattern the two humps are of slightly unequal heights,
the smaller amplitude hump preceding the eclipse,
has a small depression.  This ``splitting" of one or both humps
could be sometimes more prominent,
the periodograms of EZ Lyn often display the peak at 1/4
of orbital period (\cite{pav09j0804WD}, \cite{kat09j0804}).

\begin{figure*}
\begin{center}
\FigureFile(110mm,80mm){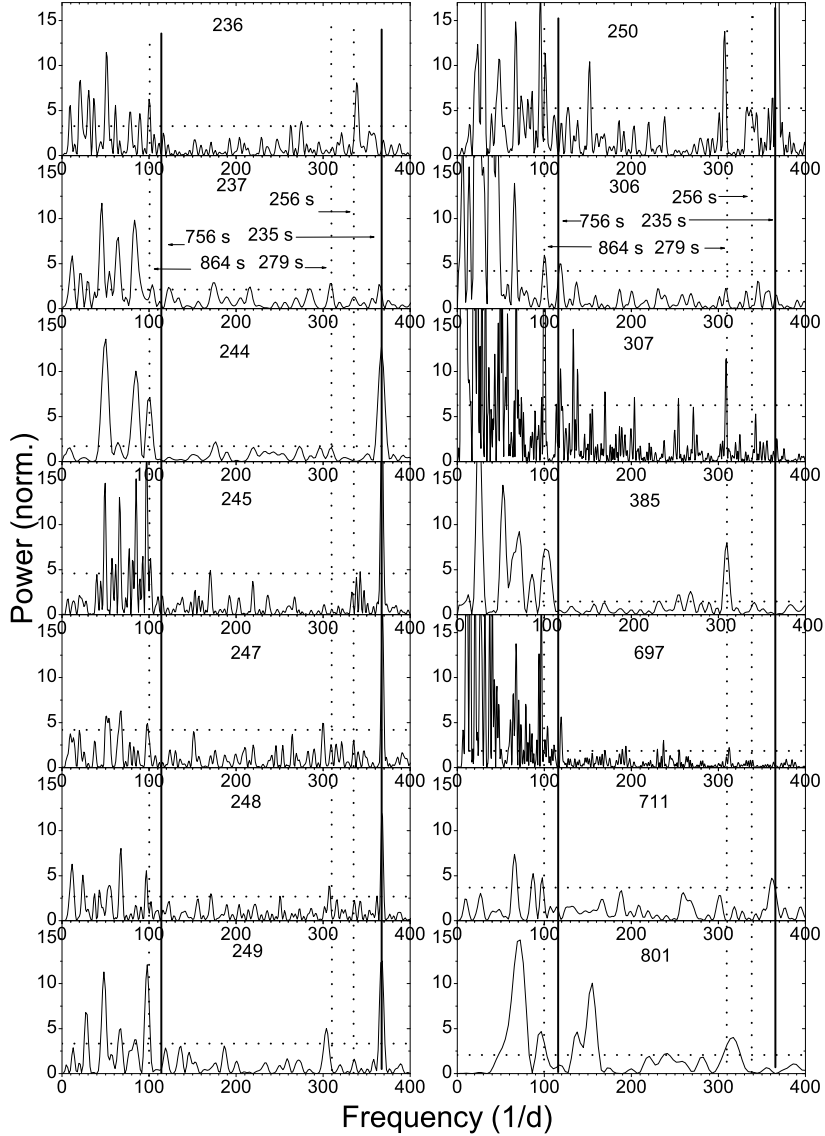}
\end{center}
\caption{The FFT for the nightly data of EZ Lyn in 2012--2014.
The numbers in each frame denote the last three digits of the JD.
The positions of dominating pulsation at 864~s and 235~s
are shown by solid lines, and those at  756~s, 279~s and 256~s
are shown by dotted lines. The horizontal line denotes the $0.1\%$ FAP level. }
\label{fig:f}
\end{figure*}

For the data of every night we subtracted the orbital modulation
with its first harmonic (that is consistent with two-humped curve)
and calculated the power spectra.
The result is given in figure \ref{fig:f}.

All the spectra contain indications to the periodic signals
both at the low and high frequency regions.
The signal around frequency
100~c/d (864~s) is surely seen
at every night (the exception is BJD 2456237, when the significance
of this signal was much lower comparing with those for other nights).
The signal around 1/4 of orbital period is visible during all nights.
The signal at 756~s being the
dominating pulsation after the 2006 outburst, which was apparently
a transient one and it is detected surely only
on BJD 2456306, 2456307 and 2456697.

Meanwhile the behavior of the high-frequency signals is more complex
and needs a detailed analysis.

The first power spectrum in this figure corresponds to the observations
obtained $\sim$2~yr since the 2010 outburst of EZ Lyn (BJD 2456236).
The 256-s pulsation that was detected by \citet{pav12ezlyn} seven months and
\citet{szk13ezlyn} twelve months after
the 2010 outburst, is the prominent and dominating one in
a high-frequency part of the periodogram.  But the next night
the spectrum changed dramatically.
The 256-s pulsation disappeared (or at least its amplitude
reduced to the noise level).  Instead one could see
the marginal signals at 279~s and 235~s.

One week later (starting from BJD 2456244) the strong 235-s signal was
detected.  It remained to be the dominating one at least
during the following week (BJD 2456244--2456250).
At the same time starting from BJD 2456248
the signal at 279~s became prominent and its amplitude has increased
up to BJD 2456248.  Again this signal dropped practically to
a noise level on BJD 2456306 and recovered on BJD 2456307.
Since this date, the 279-s signal was registered as the dominating one
in the high-frequency region.
The signal at 256~s was detected close to the 0.1$\%$ FAP only on
BJD 2456245 and BJD 2456250.

We suggest that all pulsations at the frequencies 100~c/d, 310~c/d,
338~c/d, 368~c/d
(the corresponding periods are 864~s, 279~s, 256~s and 235~s)
could be  independent modes.
However, there is an apparent linear combination of frequencies at
100~c/d ($F_{100}$), 368~c/d ($F_{368}$) and harmonic of
the orbital period $F_{\rm orb}$:
$F_{368}=3F_{100}+4F_{\rm orb}$.
So this interpretation also cannot be excluded.

The similar problem was with identification of several pulsations
in V386 Ser.  While \citet{wou04j1610} identified some pulsation
as independent one,
\citet{muk10v386ser} argued that it could be caused by
a linear combination of orbital frequency and frequency of other
known pulsation.

\section{Two-dimensional Lasso analysis}

We calculated the two-dimensional Lasso analysis for each of
14 nights separately
in the region of frequencies around 100~c/d and in
the region around high frequencies 275--400~c/d.

\begin{figure*}
\begin{center}
\FigureFile(110mm,60mm){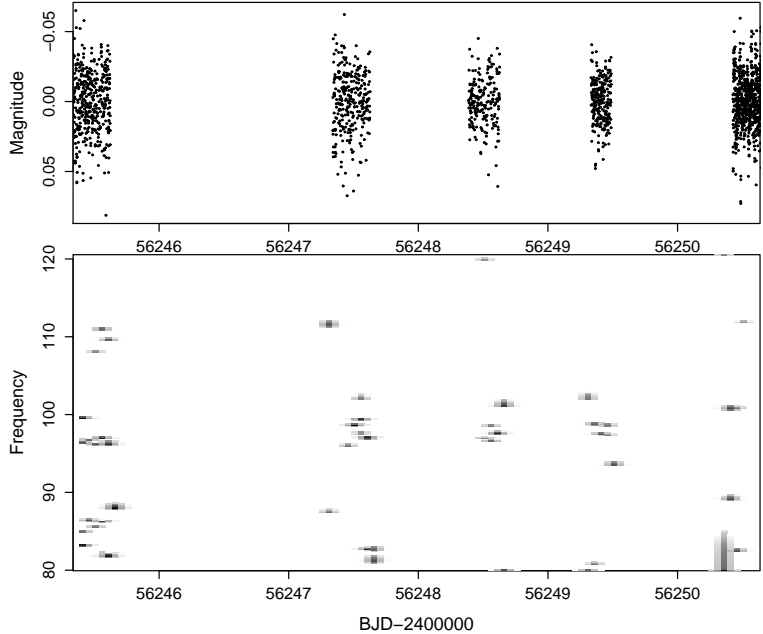}
\end{center}
\caption{Above: Light curve for BJD B2456244--2456250.
Below: Lasso analysis
 ($\log \lambda=2.5$). The width of
  the sliding window and the time step used are 0.15~d and 0.05~d,
  respectively.}
\label{fig:klasso}
\end{figure*}

\begin{figure*}
\begin{center}
\FigureFile(110mm,60mm){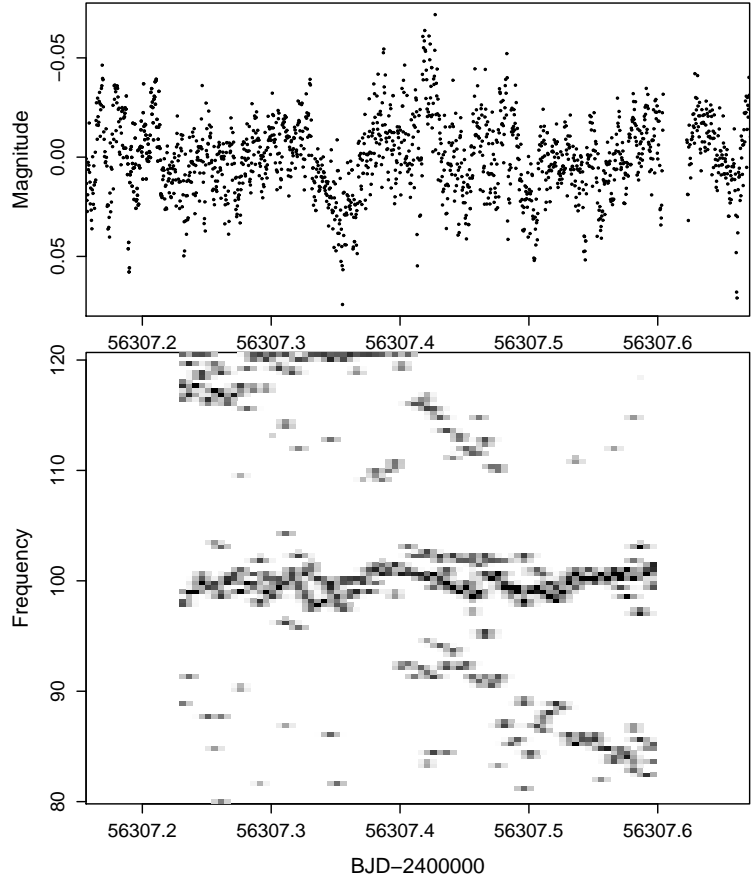}
\end{center}
\caption{Above: Light curve for BJD 2456307. Below:  Lasso analysis
($\log \lambda=-2.8$). The width of
  the sliding window and the time step used are 0.15~d and 0.005~d,
  respectively.}
\label{fig:flasso}
\end{figure*}

In figure \ref{fig:klasso} the Lasso spectra are shown
for five closely spaced nights in 2012 November.
It seems that these frequency displays
a drift both from night to night and even within a night.
More detailed behavior of this frequency is shown for the longest nightly
data on BJD 2456307 (figure \ref{fig:flasso}.
This frequency displays a smooth quasi-periodic wandering within 98--100~c/d
(or within $\sim$ 17~s) with typical time $\sim$4~hr.

\begin{figure*}
\begin{center}
\FigureFile(110mm,60mm){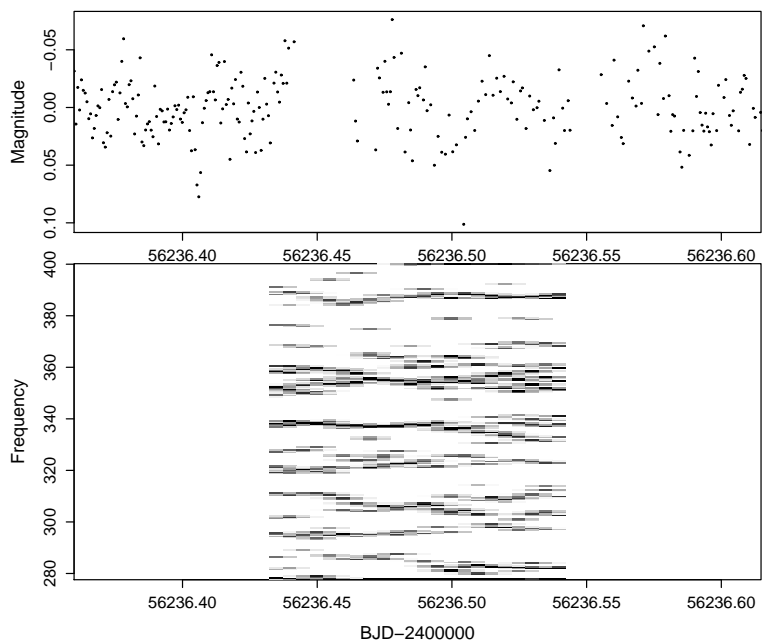}
\end{center}
\caption{Above: Light curve for BJD 2456236.
Below: Lasso analysis ($\log \lambda=-2.8$). The width of
  the sliding window and the time step used are 0.15~d and 0.005~d,
  respectively.}
\label{fig:cd1}
\end{figure*}

\begin{figure*}
\begin{center}
\FigureFile(110mm,60mm){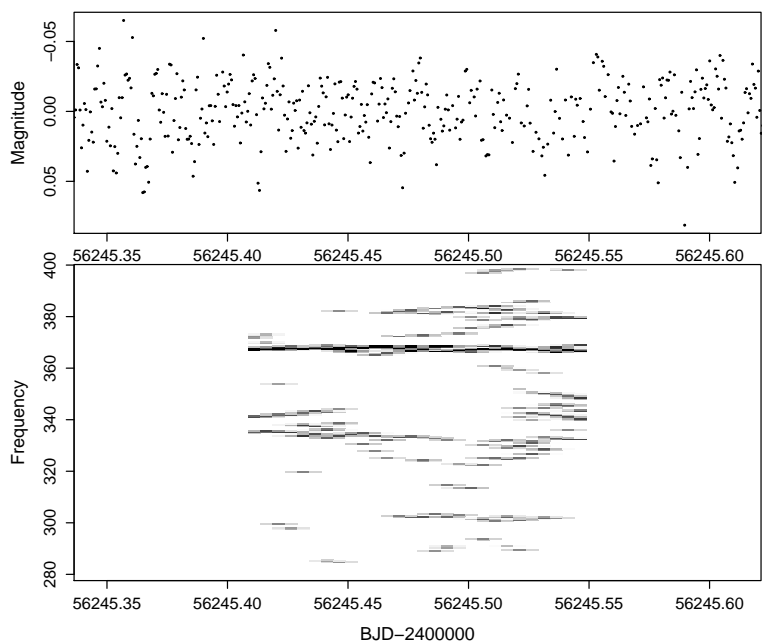}
\end{center}
\caption{Above: Light curve for BJD 2456245.
Below: Lasso analysis ($\log \lambda=-2.8$). The width of
  the sliding window and the time step used are 0.15~d and 0.005~d,
  respectively.}
\label{fig:cd4}
\end{figure*}

\begin{figure*}
\begin{center}
\FigureFile(110mm,60mm){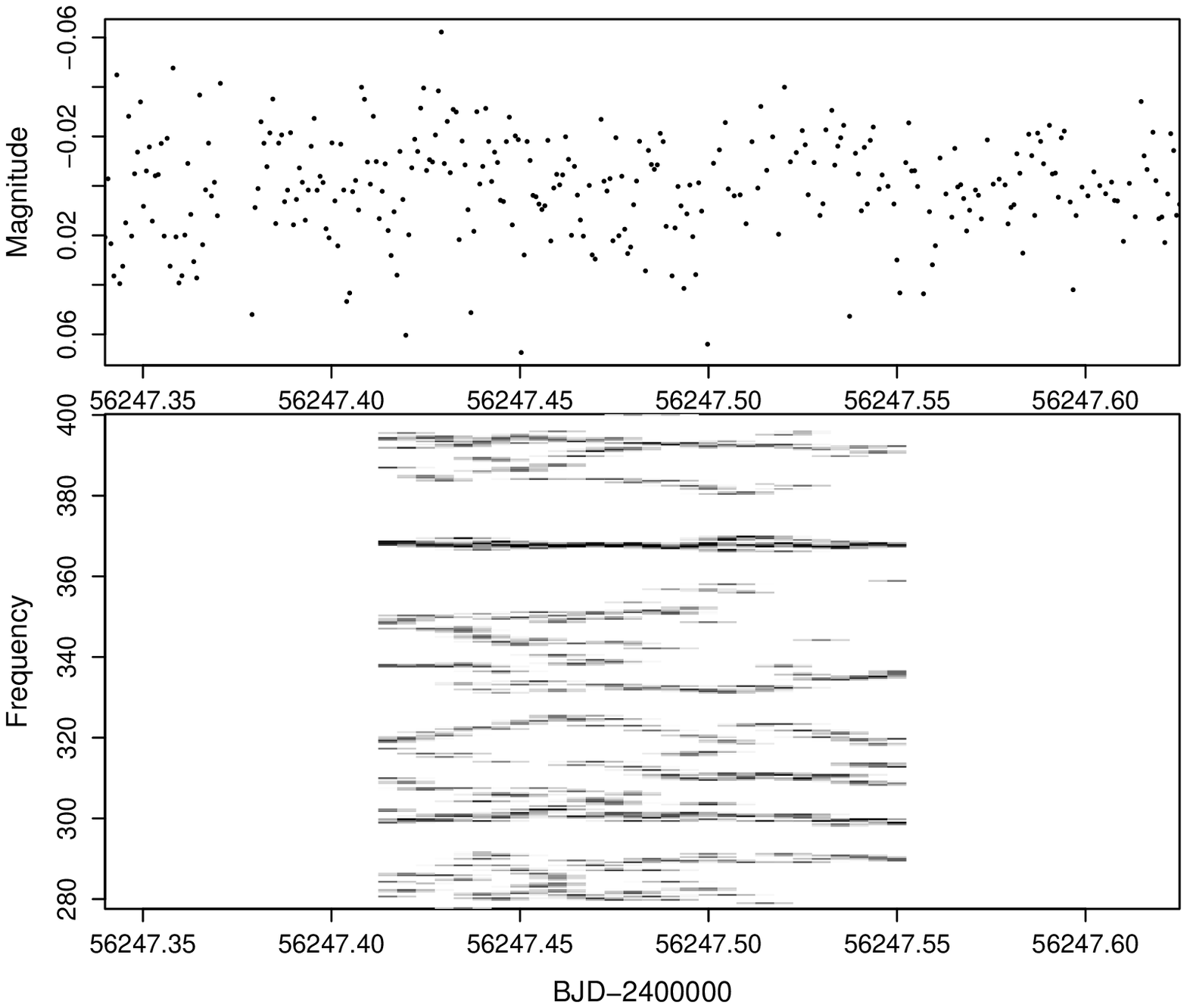}
\end{center}
\caption{Above: Light curve for BJD 2456247.
Below: Lasso analysis ($\log \lambda=-2.8$). The width of
  the sliding window and the time step used are 0.15~d and 0.005~d,
  respectively.}
\label{fig:cd5}
\end{figure*}

\begin{figure*}
\begin{center}
\FigureFile(110mm,60mm){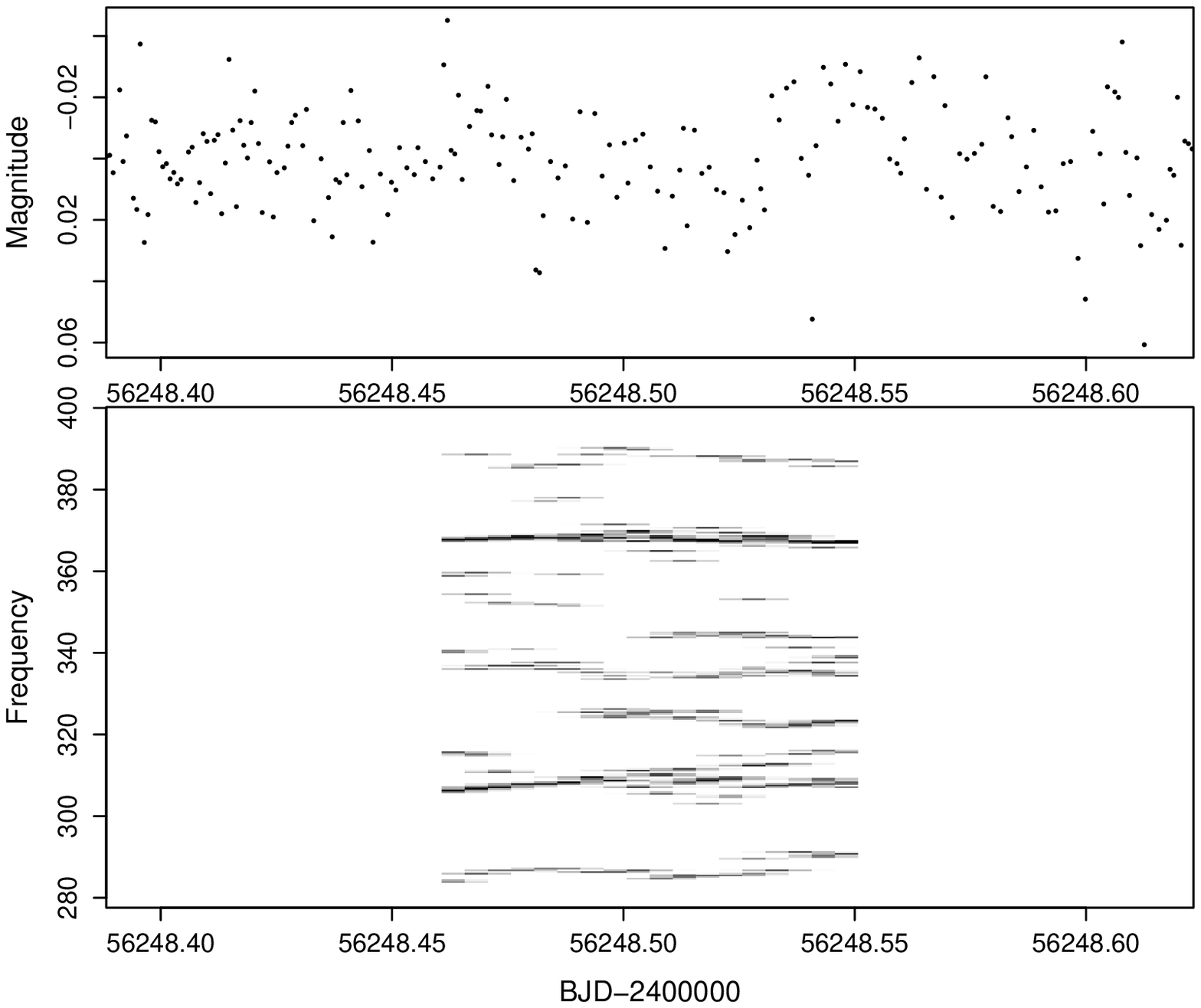}
\end{center}
\caption{Above: Light curve for BJD 2456248.
Below: Lasso analysis ($\log \lambda=-2.8$). The width of
  the sliding window and the time step used are 0.15~d and 0.005~d,
  respectively.}
\label{fig:cd6}
\end{figure*}

\begin{figure*}
\begin{center}
\FigureFile(110mm,60mm){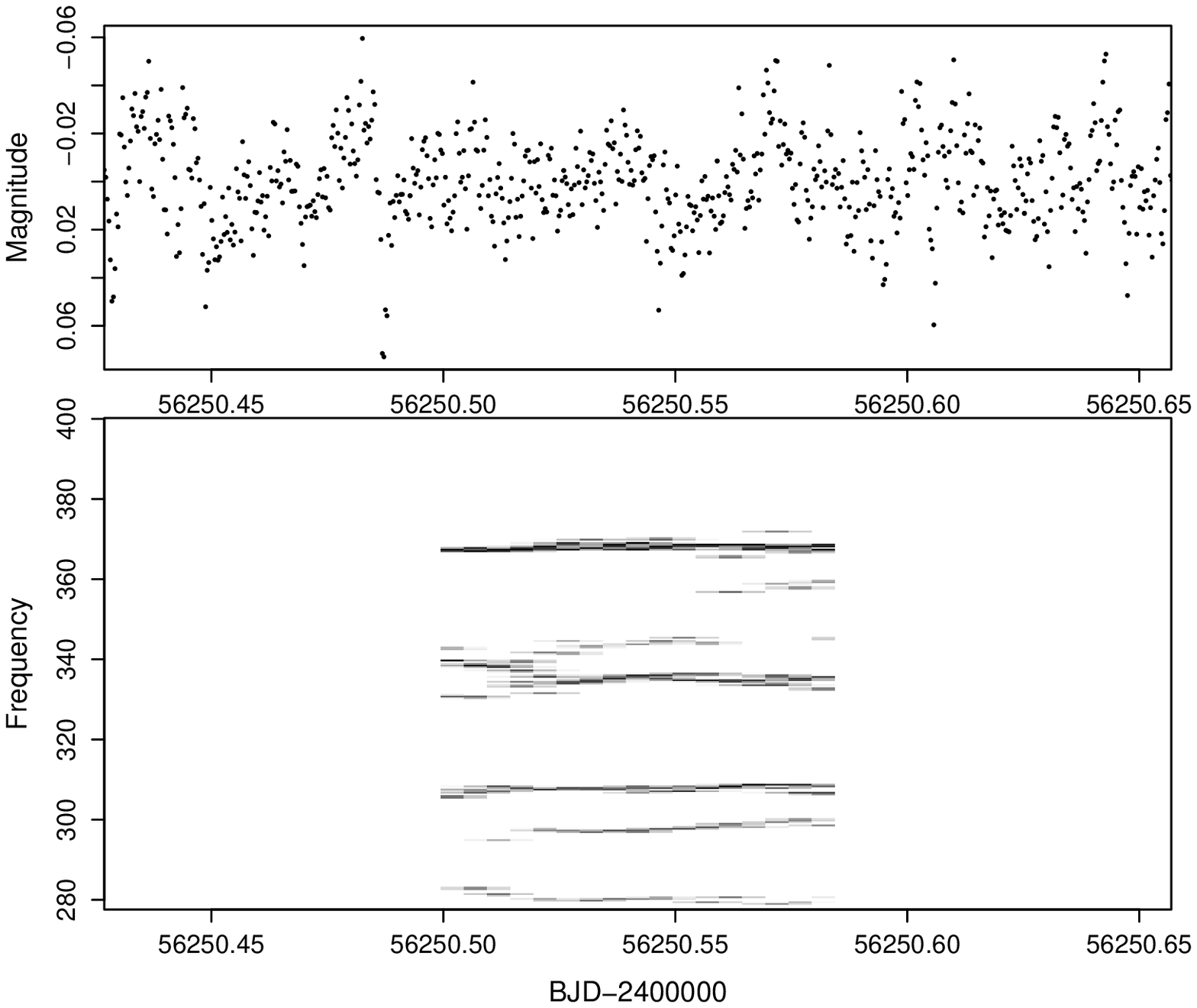}
\end{center}
\caption{Above: Light curve for BJD 2456250. Below: Lasso analysis
($\log \lambda=-2.8$). The width of
  the sliding window and the time step used are 0.15~d and 0.005~d,
  respectively.}
\label{fig:cd8}
\end{figure*}

\begin{figure*}
\begin{center}
\FigureFile(110mm,60mm){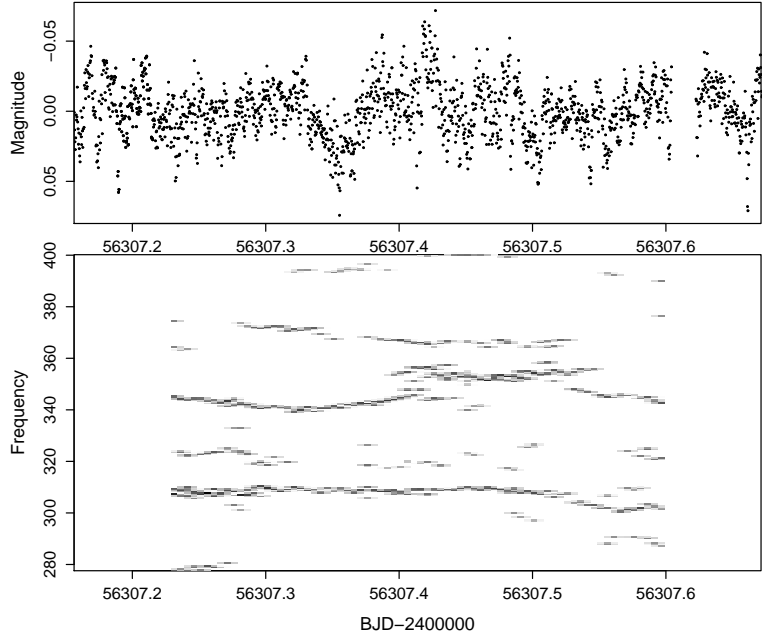}
\end{center}
\caption{Above: Light curve for BJD 2456307. Below:
Lasso analysis ($\log \lambda=-2.8$).
The width of
  the sliding window and the time step used are 0.15~d and 0.005~d,
  respectively.}
\label{fig:cd10}
\end{figure*}

\begin{figure*}
\begin{center}
\FigureFile(110mm,60mm){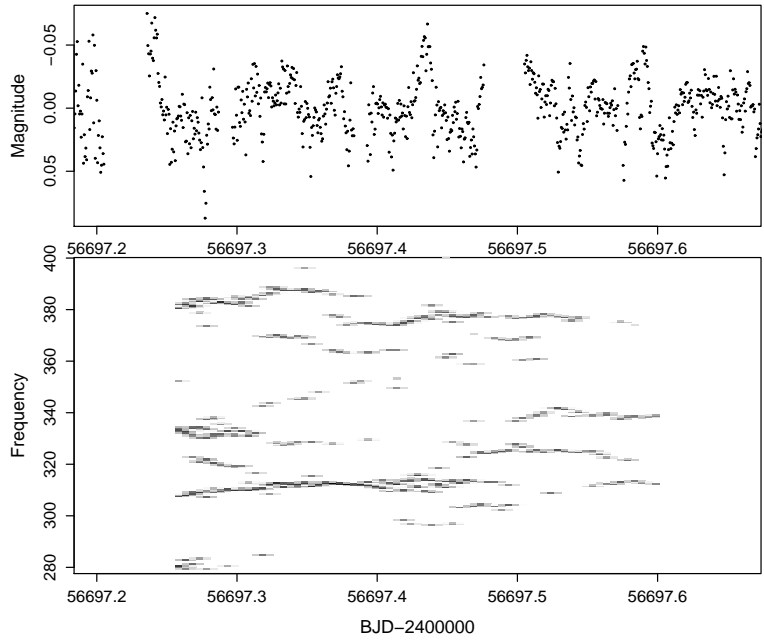}
\end{center}
\caption{Above: Light curve for BJD 2456697.
Below: Lasso analysis ($\log \lambda=-2.8$). The width of
  the sliding window and the time step used are 0.15~d and 0.005~d,
  respectively.}
\label{fig:cd12}
\end{figure*}

To study the periods in the high frequencies region, we
selected the patterns for the longest time series
(6--12 hours) on BJD 2456236, 2456245, 2456247, 2456248, 2456250,
2456307, 2456697 and presented them in the figures \ref{fig:cd1},
\ref{fig:cd4}, \ref{fig:cd5},
\ref{fig:cd6}, \ref{fig:cd8}, \ref{fig:cd10}, \ref{fig:cd12}.
After consideration of the two-dimensional power spectra
it became evident  that they give information on the behavior of
detected periods over several hours.
First, there is indication of the simultaneous co-existence
of all three periods around 279~s, 256~s and 235~s (corresponding
frequencies 310~c/d, 338~c/d and 368~c/d) for majority of nights.
Second, the domination of these periods could change within several
hours due to the non-synchronous increase or decrease of
their amplitudes and some drift of period.

The data on BJD 2456236 show that the amplitude of the signal
at the dominating frequency
around 338~c/d was higher during the initial several hours.
There is no clear indication of a signal at 368~c/d and
there is some indication of a weak signal at 310~c/d
of variable amplitude and phase.
The signal at the dominating frequency
around 368~c/d looks stable in its amplitude and period on
BJD 2456245. There is a weaker but prominent signal
at 338~c/d, which is stronger  during the first hours of the observation.
The frequency of this signal displays some decrease with time.
The signal at frequency around 368~c/d for BJD 2456247, 2456248
and 2456250 remains to be the dominating one.
The signal at frequency 338~c/d is rather marginal on
BJD 2456247 and BJD 2456248, but is strong on BJD 2456250.
The signal at frequency 310~c/d is strong for all these nights.

The longest 12-hr data set on BJD 2456307 revealed a signal
at 310~c/d of variable
amplitude and near-constant period that produces the dominating peak
on the FFT periodogram.  The most unusual thing is the behavior of
a rather strong signal that varies between frequencies
340~c/d--352~c/d (this variation corresponds to 9~s in period)
quasi-periodically on a typical time scale of $\sim$7 hr.
This causes a ``smeared" peak on the FFT periodogram.
We believe this signal is related to those at frequency 338~c/d.
Using all the data set, this signal varied by more than 9~s
in period.
There was no strong indication to the signal at 368~c/d.

The two-dimensional power spectrum on BJD 2456697 only
shows a signal at 310~c/d with highly variable amplitude and increasing
frequency from $\sim$305~c/d to $\sim$325~c/d that
corresponds to the drift of 17.5~s in period in $\sim$5 hours.

\section{Summary}

The analysis of 14 periodograms of EZ Lyn for the data spaced
over 565~d in 2012--2014 (2--3.5 yr after 2010 outburst) yielded
the existence of several signals and their evolution.
These are signals at the frequencies 100~c/d, 310~c/d, 338~c/d, 368~c/d
(the corresponding periods are 864~s, 279~s, 256~s and 235~s).
We believe that these signals arise from non-radial pulsations
in the accreting white dwarf.

The transient appearance of pulsation at 846~s was detected
during the first stay of EZ Lyn in the instability strip
(\cite{pav09j0804WD}, \cite{pav12ezlyn}), and the pulsation
at 256~s was the pulsation when EZ Lyn entered the instability strip
the second time.
We detected the pulsations at 279 and 235~s for the first time.
We suggest that all pulsations could be the independent modes,
however a linear combination of frequencies at 100~c/d, 368~c/d
and harmonic of the orbital period also cannot be excluded.

Our knowledge of a fast change in pulsation period and amplitude
before this work had been rather limited. Recently \citet{uth12j1457bwscl}
reported that pulsations in accreting white dwarfs BW Scl and
SDSS J145758.21$+$514807.9 change in frequency by a few percent
on a timescale of weeks or less; \citet{cho13} found that in GW Lib a
strong non-sinusoidal pulsation with a 19 minute period varied slightly
in frequency over the six nights of observations.

Here we showed that the pulsations
of EZ Lyn change in frequency by 2--6 percents on a time scale
of hours and their
amplitude may change on the same time scale.
We found that Lasso is very powerful in detecting multiple
frequencies and their variations in the pulsating white dwarfs.

\medskip

We thank the anonymous referee whose
comments greatly improved the paper.
This work was supported by the Grant-in-Aid Initiative for High-Dimensional
Data-Driven Science through Deepening of Sparse Modeling from
the Ministry of Education,
Culture, Sports, Science and Technology (MEXT) of Japan.
The work  is partially performed according to the Russian Government Program of
Competitive Growth of Kazan Federal University; EP thanks to V. Malanushenko
for valuable comments;
SA, BI, GA thank to TUBITAK and KFU for partial support in using RTT150 (Russian-Turkish 1.5-m telescope in Antalya).

\newcommand{\noop}[1]{}


\begin{thebibliography}{}

\bibitem[{Araujo-Betancor} et~al.(2005)]{ara05v455and}
  {Araujo-Betancor}, S., {et~al.}\ 2005, A\&A, 430, 629

\bibitem[{Chote}, {Sullivan}(2013)]{cho13}
  {Chote}, P., \& {Sullivan}, D.~J.\ 2013, in ASP\ Conf.\ Ser.\ 469, 18th European
  White Dwarf Workshop, ed. J. Krzesinski, G. Stachowski, P. Moskalik, \& K. Bajan (San
  Francisco: ASP), p.~337

\bibitem[Fernie(1989)]{fer89error}
  Fernie, J.~D.\ 1989, PASP, 101, 225

\bibitem[{Fontaine}, {Brassard}(2008)]{fon08pulsWDreview}
  {Fontaine}, G., \& {Brassard}, P.\ 2008, PASP, 120, 1043

\bibitem[{G{\"a}nsicke} et~al.(2006)]{gan06j1339}
  {G{\"a}nsicke}, B.~T., {et~al.}\ 2006, MNRAS, 365, 969

\bibitem[Howell et~al.(1995)]{how95TOAD}
  Howell, S.~B., Szkody, P., \& Cannizzo, J.~K.\ 1995, ApJ, 439, 337

\bibitem[{Kato} et~al.(2014)]{Pdot5}
  {Kato}, T., {et~al.}\ 2014, PASJ, 66, 30

\bibitem[{Kato}, {Maehara}(2013)]{kat13j1924}
  {Kato}, T., \& {Maehara}, H.\ 2013, PASJ, 65, 76

\bibitem[{Kato} et~al.(2012)]{Pdot3}
  {Kato}, T., {et~al.}\ 2012, PASJ, 64, 21

\bibitem[{Kato} et~al.(2010)]{Pdot2}
  {Kato}, T., {et~al.}\ 2010, PASJ, 62, 1525

\bibitem[{Kato}, {Osaki}(2013)]{kat13j1939v585lyrv516lyr}
  {Kato}, T., \& {Osaki}, Y.\ 2013, PASJ, 65, 97

\bibitem[{Kato} et~al.(2009)]{kat09j0804}
  {Kato}, T., {et~al.}\ 2009, PASJ, 61, 601

\bibitem[{Kato}, {Uemura}(2012)]{kat12perlasso}
  {Kato}, T., \& {Uemura}, M.\ 2012, PASJ, 64, 122

\bibitem[{Knigge}(2006)]{kni06CVsecondary}
  {Knigge}, C.\ 2006, MNRAS, 373, 484

\bibitem[Kolb, Baraffe(1999)]{kol99CVperiodminimum}
  Kolb, U., \& Baraffe, I.\ 1999, MNRAS, 309, 1034

\bibitem[{Mukadam} et~al.(2007)]{muk07j0745j0919}
  {Mukadam}, A.~S., {G{\"a}nsicke}, B.~T., {Szkody}, P., {Aungwerojwit}, A.,
  {Howell}, S.~B., {Fraser}, O.~J., \& {Silvestri}, N.~M.\ 2007, ApJ, 667, 433

\bibitem[{Mukadam} et~al.(2010)]{muk10v386ser}
  {Mukadam}, A.~S., {et~al.}\ 2010, ApJ, 714, 1702

\bibitem[{Nilsson} et~al.(2006)]{nil06SDSSCVpuls}
  {Nilsson}, R., {Uthas}, H., {Ytre-Eide}, M., {Solheim}, J.-E., \& {Warner},
  B.\ 2006, MNRAS, 370, L56

\bibitem[{Ohshima} et~al.(2014)]{ohs14eruma}
  {Ohshima}, T., {et~al.}\ 2014, PASJ, in press (arXiv/1402.5747)

\bibitem[{Osaki}, {Kato}(2013)]{osa13v344lyrv1504cyg}
  {Osaki}, Y., \& {Kato}, T.\ 2013, PASJ, 65, 95

\bibitem[{Patterson} et~al.(2005)]{pat05j1255}
  {Patterson}, J., {Thorstensen}, J.~R., \& {Kemp}, J.\ 2005, PASP, 117, 427

\bibitem[{Patterson} et~al.(2008)]{pat08j1507}
  {Patterson}, J., {Thorstensen}, J.~R., \& {Knigge}, C.\ 2008, PASP, 120, 510

\bibitem[{Pavlenko}(2009)]{pav09j0804WD}
  {Pavlenko}, E.\ 2009, J.\ of\ Physics\ Conference\ Series, 172, 012071

\bibitem[{Pavlenko} et~al.(2012)]{pav12ezlyn}
  {Pavlenko}, E., {et~al.}\ 2012, Mem.\ Soc.\ Astron.\ Ital., 83, 520

\bibitem[{Pavlenko} et~al.(2007)]{pav07j0804}
  {Pavlenko}, E., {et~al.}\ 2007, in ASP\ Conf.\ Ser.\ 372, 15th European
  Workshop on White Dwarfs, ed. R. Napiwotzki, \& M.~R. Burleigh (San
  Francisco: ASP), p.~511

\bibitem[{Pel't}(1980)]{pel80freqanalysis}
  {Pel't}, {Ya}.\ 1980, Frequency analysis of astronomical time series.
  (Tallin: Valgus)

\bibitem[{Scargle}(1982)]{sca82}
  {Scargle}, J.~D.\ 1982, ApJ, 263, 835

\bibitem[{Shears} et~al.(2007)]{she07j0804}
  {Shears}, J., {Klingenberg}, G., \& {de Ponthiere}, P.\ 2007, J.\ Br.\
  Astron.\ Assoc., 117, 331

\bibitem[Stellingwerf(1978)]{PDM}
  Stellingwerf, R.~F.\ 1978, ApJ, 224, 953

\bibitem[{Szkody} et~al.(2006)]{szk06SDSSCV5}
  {Szkody}, P., {et~al.}\ 2006, AJ, 131, 973

\bibitem[{Szkody} et~al.(2010)]{szk10CVWDpuls}
  {Szkody}, P., {et~al.}\ 2010, ApJ, 710, 64

\bibitem[{Szkody} et~al.(2013)]{szk13ezlyn}
  {Szkody}, P., {Mukadam}, A.~S., {Sion}, E.~M., {Gansicke}, B.~T., {Henden},
  A., \& {Townsley}, D.\ 2013, AJ, 145, 121

\bibitem[{Tibshirani}(1996)]{lasso}
  {Tibshirani}, R.\ 1996, J. R. Statistical Soc. Ser. B, 58, 267

\bibitem[{Uthas} et~al.(2012)]{uth12j1457bwscl}
  {Uthas}, H., {et~al.}\ 2012, MNRAS, 420, 379

\bibitem[{Vanlandingham} et~al.(2005)]{van05pqand}
  {Vanlandingham}, K.~M., {Schwarz}, G.~J., \& {Howell}, S.~B.\ 2005, PASP,
  117, 928

\bibitem[{Warner}, {van Zyl}(1998)]{war98gwlibproc}
  {Warner}, B., \& {van Zyl}, L.\ 1998, in 185, New Eyes to See Inside the Sun
  and Stars, ed. F.-L. {Deubner}, J. {Christensen-Dalsgaard}, \& D. {Kurtz}
  (Berlin: Springer), p.~321

\bibitem[{Winget}, {Kepler}(2008)]{win08pulsWDreview}
  {Winget}, D.~E., \& {Kepler}, S.~O.\ 2008, ARA\&A, 46, 157

\bibitem[{Woudt}, {Warner}(2004)]{wou04j1610}
  {Woudt}, P.~A., \& {Warner}, B.\ 2004, MNRAS, 348, 599

\bibitem[{Zharikov} et~al.(2008)]{zha08j0804}
  {Zharikov}, S.~V., {et~al.}\ 2008, A\&A, 486, 505

\end{thebibliography}
\end{document}